\def\be{\begin{equation}}
\def\ee{\end{equation}}
\def\te{\end{equation}}
\def\bea{\begin{eqnarray}}
\def\ba{\begin{eqnarray}}

\def\ta{\end{eqnarray}}
\def\tea{\end{eqnarray}}

\def\ben{\begin{enumerate}}
\def\een{\end{enumerate}}

\def\m{\mu}

\def\bfr{{\bf r}}

\documentclass[12pt]{iopart}
\begin{document}

\title{Gravitational Decoherence, Alternative Quantum Theories and Semiclassical Gravity}

\author{B. L. Hu}
\address{Maryland Center for Fundamental Physics and Joint Quantum Institute,\\ University of
Maryland, College Park, Maryland 20742-4111 U.S.A.}
\ead{blhu@umd.edu}

\date{February 26, 2014}

\begin{abstract}
In this report we discuss three aspects: 1) Semiclassical gravity theory (SCG): 4 levels of theories describing the interaction of  quantum matter with classical gravity; 2) Alternative Quantum Theories: Discerning those which are  derivable from general relativity (GR) plus quantum field theory (QFT) from those which are not; 3) Gravitational Decoherence:  derivation of a master equation  and examination of the assumptions which led to the claims of  observational possibilities. We list three sets of corresponding problems worthy of pursuit: a) Newton-Schr\"odinger Equations in relation to SCG; b) Master equation of gravity-induced effects serving as discriminator of 2); and  c) Role of gravity in macroscopic quantum phenomena.
\end{abstract}

\section{Introduction}

There is general agreement that general relativity (GR) is an excellent theory describing the large scale structures of spacetime and quantum field theory (QFT) a highly successful theory for matter down to the verifiable subnuclear levels. Yet, it is equally well-accepted that intrinsic contradictions between general relativity and quantum theories exist. There are many serious efforts to  reconcile or unify them in the search of \textit{a theory for the microscopic structures of spacetime}, which is what \textbf{quantum gravity} (QG) entails -- and carries no other meaning, specifically not quantizing general relativity (see, e.g., \cite{E/QG})-- but it is fair to say to date no one school can show definitive and complete success in this goal.

\subsection{Semiclassical Gravity}

A modest yet no less productive attempt is to place these two theories together: Q $\oplus$ G, not Q $\otimes$ G,  which we don't yet quite understand how to do -- GR being a classical theory for the  macroscopic realm while QFT a quantum theory for the microscopic world, and see what discrepancies  this union may reveal, such as in their mathematical structures, and what new physical insights we may gain. This was the  goal set in \textbf{quantum field theory in curved spacetime} (QFTCST) \cite{BirDav,Wald,ParTom} which began in the late 60's with cosmological particle creation studies \cite{Par69} and epitomized in  Hawking's 1974 \cite{Hawking} discovery of black hole radiance. Focused efforts in seeking ways to regularize or renormalize the stress energy tensor of quantum fields made it possible to tackle the so-called `backreaction problem' \cite{cosbkrn,bhbkrn} in finding how quantum matter fields affect the dynamics of spacetime.

Solving the backreaction problem is at the core of \textbf{semiclassical gravity} theory (SCG) \cite{scg}  developed in the 80's based on the \textit{semiclassical Einstein equation }(SCE).  Discovery in the 90s of a lawful place  for the fluctuations of quantum fields promoted this to the \textit{Einstein-Langevin equation} \cite{ELE} which enables one to solve for the induced metric fluctuations (Wheeler's poetic `spacetime foam'). This ushered in a new theory known as \textbf{stochastic gravity} \cite{stograCQG,HuVerLivRev}. Both theories have since been developed extensively and applied to strong field situations such as structure formation in the early universe and black hole fluctuation and backreaction issues.

As a summary remark, the validity of semiclassical gravity in the form first proposed by M\/oller and Rosenfeld \cite{MolRos} in the 60's is often raised by authors of Newton-Schr\"odinger equation, citing the arguments by Page and Geilker \cite{PagGei}, Eppley and Hannah \cite{EppHan}\footnote{Since many of these theories which combine classical gravity with quantum mechanics are now also referred to as semiclassical gravity by many practitioners,  to distinguish from them (which we see below are important) we will call theories based on quantum field theory (second quantized, permitting particle creation) in curved spacetime, which is a well-established form of GR+QFT,  Relativistic Semi-Classical Gravity (RSCG).}.  Leaving aside the question of whether gravity should be quantized, which had seen much broader and deeper discussions since then, the internal consistency of relativistic semiclassical gravity by itself had since been investigated further and there are better responses to the challenges posed in the early 80s (read e.g. papers by Kibble \& Randjbar-Daemi and Duff in \cite{QG2}). We refer to two substantive papers, one by Flanagan and Wald \cite{FlaWal} on semiclassical gravity, the other by Hu, Roura and Verdaguer \cite{HuRouVer}, which also considered the role of the induced metric fluctuations in the criteria.

\subsection{Alternative Quantum Theories (AQT)}

General relativists following this vein have probed the interplay between gravity and quantum largely from the angle of how quantum matter affects spacetime (Q $\rightarrow$ G). Asking the question in the other direction (G$\rightarrow$ Q), namely, how gravity could have an effect on quantum phenomena, has been going on for just as long (e.g., \cite{Karol}) mainly by quantum foundation theorists.  The foremost issue is why  macroscopic objects are found sharply localized in space (their wave functions ``collapsed" on definite locales) while those of microscopic objects extend over space. This contradiction is captured in the celebrated Cat of Schr\"odinger \footnote{Note that `cat-states' have been found for atoms whereas entangled `dead' and `alive' state for real cats, which are a little bigger than atoms, have not. The difference between micro and macro objects is crucial insofar as their quantum behaviors are concerned. A missing basic ingredient is nonequilibrium statistical mechanics which helps us interpolate between micro/few body effects and macro/many-body phenomena. Here lies the importance of macroscopic quantum phenomena (MQP), a subject hitherto largely overlooked by theorists but I feel is essential in understanding why cat-states can never be found for real cats. }. One can very coarsely place these theories in three groups: The Girahdi-Remini-Weber (GRW)- Pearle  models \cite{GRWP} of  continuous spontaneous localization (CSL),  the Diosi-Penrose theories \cite{Diosi,Penrose} invoking gravitational decoherence,  and the recent  trace dynamics theory of Adler \cite{AdlerCUP} which attempts to provide a sub-stratum theory from which quantum mechanics emerges. A nice description of these theories can be found in a recent review by Bassi et al \cite{BassiRMP}.
%\footnote{We can name  four levels of theories of quantum matter interacting with classical gravity, all different and yet all called semiclassical gravity theories. At Level 0 are the NS Eqs.  Level 1 has first quantized quantum field, e.g., Einstein-Klein-Gordon Eq. Level 2 has  second quantized fields, e.g., Semiclassical Einstein Eq, which we call relativistic semiclassical gravity (RSCG) and Level 3 has second quantized fields plus their fluctuations, e.g., Einstein-Langevin equation, which we call stochastic gravity.}

%\bigskip \noindent\textbf{

\subsection{Gravitational Decoherence}

One important process where the interplay of gravity with quantum manifest is gravitational decoherence  -- the mechanism where the quantum coherence of a particle is diminished due to its interaction with an environment, in this case provided by the gravitational field. (The differences between quantum, intrinsic and gravitational decoherence are explained in the Introduction of \cite{AHmasteq}. See also \cite{AHdecQG,AHintDec}).
Gravitational decoherence is invoked in an important class of AQTs, that by the name of Diosi-Penrose theories.  We want to find out the special features of gravitational decoherence, such as the decoherence rate and the associated basis and how gravitational decoherence differs from decoherence by a non-gravitational environment. To perform quantitative analysis of this effect one needs a  master equation which has not been derived from first principles until only recently (see papers by Blencowe and by Ananstopoulos \& Hu  below). Earlier equations have been reasoned out rather than derived from known microscopic physics. The `reasoning out' process admits inputs based on phenomenological arguments according to the proponents' particular wishes.
Whether gravity can be an effective source of decoherence is a reasonable  motivation to work out a master equation for such analysis. More of this in Sec. 4.

\subsection{Experimental Possibilities}

While the early universe and black holes are the natural arena where strong field quantum processes play out, which necessitate QFTCST and RSCG, the weak field and nonrelativistic limits are certainly more within the reach of what laboratory experiments can measure. Rapidly improving precision levels of observational possibilities in molecule interferometry, optomechanics and mirco-trap experiments \cite{GravDecExpts} are pushing this closer to reality. While tests of AQTs are understandably high on the agenda of the proponents of such theories, the more modest yet equally if not more important task is to ``put SCG to test" (note SCG here is used in the weak field nonrelativistic context) as exemplified by Chen's group's \cite{ChenNS} recent application of the many particle NS equation to macroscopic objects and  estimating the difference in their predictions from that of standard quantum mechanics (see below).

Designs of experimental setups and tests are already underway which involve theorists working in  QFTCST. Witness the increased activities in precision measurement of Casimir effect, dynamical Casimir effect, for understanding the vacuum energy modified by boundaries, and the parametric amplification thereof, which is the mechanism underlying cosmological particle creation. Jets and bursts of atoms from the controlled collapse of a BEC (``Bosenova") can be used to understand particle creation in inflationary universe \cite{CHbosenova}, an example of laboratory cosmology. Accelerating atoms for testing the Unruh effect and finding analogs of Hawking effect in BECs and moving mirrors are being pursued. These activities belong to the realm of a new field called analog gravity \cite{AnalogG}. Theoretical results obtained in the 70s are now being improved on  and applied to designs for possible experimental verifications. In our opinion, tests related to RSCG will come next. Now is a good time for researchers in the 80s (on RSCG) and 90s (on semiclassical stochastic gravity) to join force with experimental activities so their expertise can enrich our understanding of the interplay between quantum matter and classical gravity.

\subsection{Theoretical preparatory work for observing gravitational decoherence and testing AQTs}

AQTs have been in existence for 3 decades, but the tests of such theories are put in practice only recently  thanks to the increasing precision  required of the measurements of these gravity-induced effects (see, e.g., \cite{GravDecExpts}). Indications of the timeliness for these investigations are seen also in concentrated recent activities, e.g., 6 papers of substance have appeared in the last 6 months. We list five of them here with a short description because we will refer to all of them in this report.  Noteworthy also is Adler's recent work incorporating gravity into his trace dynamics theory \cite{AdlerTDG}. Increased effort in theoretical investigations of NS equations and gravitational decoherence are needed to better prepare the ground for measurement possibilities.\\

% the claims of observation close to theoretical predicted level -- one needs to first show the intrinsic relevance of SCG to AQTs and the need for bringing SCG closer to experiments.

1) Dyson's 2012 essay, \cite{Dyson12} ``Is a graviton detectable?" will be useful for our discussions of gravitational decoherence, e.g., whether a thermal bath of gravitons is for real.

2) Review by Bassi et al \cite{BassiRMP} has a lucid summary of alternative quantum theories and a clear description of the experimental results in progress (See Table 1,  also \cite{RomIsa}.)

3) Chen's group \cite{ChenNS} (\textbf{Version C}.) applies the NS equation for N particles proposed by Diosi et al to macroscopic objects, assesses the differences between the predictions of NS and ordinary QM and  their experimental observability. We will show in Sec. 3 the basic differences between the many-particle NS equations and that obtained from taking the nonrelativistic limit of SCE equation. The single or many particle NS equations are not derivablem from GR + QFT \cite{AH-NS1}.

4) Anastopoulos and Hu \cite{AHmasteq} derived a master equation for gravitational decoherence based solely on GR+QFT. We call this theory \textbf{Version A}. The procedure and results of this paper are summarized in Section 4.

5) Blencowe \cite{Blen13} derived a master equation using the influence functional method and made claims to the effect that gravitational decoherence is strong enough to soon be within observational range. (We call his master equation \textbf{Version B}.)  We have reservations in his claim which invokes two assumptions:  a thermal bath of gravitons and the increase of decoherence strength with mass. We will comment on these two assumptions in Section 7. \\

%For example, the magnitude of gravitational decoherence derived from our master equation is much weaker than that of Blencowe's. We have doubts in two assumptions made there

In all, our present work attempts to address the core issues in the interplay between gravity and quantum, places the Newton-Schr\"odinger equation in the context of relativistic semiclassical gravity and  explores the theoretical base for observational possibilities of gravitational decoherence.
%and investigates the effects of quantum fluctuations as source of noise in AQTs.
Our hope is that the theoretical structure developed based on known physics from general relativity and quantum field theory and the results obtained here can provide a definitive standard for all alternative quantum theories to compare in their reasonings and predictions. The new results reported here on the master equation for gravitational decoherence and the NS equation in relation to SCG are based on two recent papers \cite{AHmasteq,AH-NS}

%%%%%%%%%%%%%%%%%%%%%%%%%%%%%%%%%%%%%%%%%%%%%%%%%%%%%%%%%%%%%%%%%%%%%%%%%%%%%%%%
%\vskip .5cm \centerline{\Large {\bf Part I.  Newton-Schr\"odinger Equations of different forms  }}
%\centerline{\Large {\bf   in relation to Semiclassical Einstein Equation}}
%\vskip .3cm
%%%%%%%%%%%%%%%%%%%%%%%%%%%%%%%%%%%%%%%%%%%%%%%%%%%%%%%%%%%%%%%%%%%%%%%%%%%%%%%

\section{Quantum matter interacting with classical gravity: 4 Levels of inquiry}

%Quantum particle: nonrelativistic vs relativistic. Nonrelativistic QM: Schr\"odinger. Relativistic matter described by quantum field theory. Gravity: strong field (in stars and black holes, the early universe) vs weak field. Weak field limit called Newtonian, Quantized weak field = graviton but this is not about quantum gravity, a theory for the microscopic structures of spacetime \cite{E/QGrav}.

Newton-Schr\"odinger (NS) equations describe the motion of nonrelativistic quantum particle(s) in a weak gravitational field potential. There are many forms,  justified with different rationales, originated from different motivations.  Some  are in conflict with general relativity as they are not the nonrelativistic limit of relativistic semiclassical gravity (RSCG).
%This is analyzed below. %NS eqn has been written down for single particles and only recently for many particles.
From their formal appearance one may view the NS equation as the nonrelativistic limit of the Einstein-Klein-Gordon equation, but NS in structure is closer to the Hartree-Fock equations. Yet, despite the similarity in structure between the NS eqn and Hartree-Fock equations, there are differences between gravito- and electro-statics) in the self energy for a many body system.
These issues need be clarified so that one knows what his/her results of calculation pertain to in relation to other theories and in comparison with experiments.

\subsection{Level 0: Newton-Schr\"odinger / Schr\"odinger-Poisson Eqs: non-relativistic quantum particle in a weak gravitational field}

This is the arena where most of the activities in finding / showing the overt / hidden effects of gravity on quantum mechanics takes place and proposals of alternative quantum theories reside. Because it is for non-relativistic particles in a weak gravitational field, this is also the domain where laboratory experiments are carried out.  Most researchers working in the 70s  on quantum field theory in curved spacetime (QFTCST) and in the 80s working on (relativistic) semiclassical gravity (RSCG) (see below) have largely ignored this level of activities going on at the same time, albeit sparingly, as much as researchers working on quantum foundational issues largely were unaware of developments in RSCG in the last four decades. The former group's attention was focused on quantum effects in strong gravitational fields as in the early universe and in black holes while the latter group was focusing on proposing alternative quantum theories (AQTs) and their experimental verification possibilities. The latter group's definition of SCG is largely within the realm of Level 0.

So what are the problems of interest at this Ground Level? One can start with an almost textbook-like example of calculating  how the dispersion of a Gaussian wave packet with initial spread  $a_0$ of a massive (m) particle according to the Schr\"odinger equation
$ i\hbar \partial \psi/ \partial t = - \frac{\hbar^2}{2m} \nabla^2 \psi -m V \psi$ when it moves in a gravitational potential V sourced by its own wavefunction  $\nabla^2 V = 4 \pi Gm |\psi|^2$. This set of Newton-Schr\"odinger equation is imbued with a tension illustrated in this simple example between the natural quantum  dispersion of a wavepacket against the gravitational collapse due to its own mass. The rather low critical mass obtained by Salzman and Carlip \cite{Carlip,SalCar} in 2006 was contested by Giulini and Grossrdtin \cite{Giulini} in 2010.  By introducing a length scale $\ell$ the SN equation can be written in terms of a dimensionless coupling constant, $K =  2 (\ell/\ell_P)(m/m_P)^3$ where $\ell_P, m_P$ are the Planck length and mass respectively.  Giulini and Grossrdtin  showed that inhibition of the dispersion becomes significant when the dimensionless coupling constant K becomes of order unity. Their conclusion (quoting from \cite{BassiRMP})``.. leads to an important inference: The models of Karolyhazy, Diosi, and Penrose all agree that if the width of the quantum state associated with an object of mass m becomes greater than of the order $\hbar^2/(Gm^3)$, the quantum-to classical transition sets in. For the experimentally interesting $a= 0.5 \m m$ this gives $m$ of about $10^9$ amu." This estimate puts the actual measurement of this effect beyond today's experimental capability but  the intellectual challenge and excitement towards realization of this goal are certainly growing. More examples to expound the interplay between  effects of quantum dispersion and gravitational pull in wave packets and more complex systems can be explored at the next level (Level 1) with the help of quantum field theory in curved spacetime techniques.

%explore the theoretical basis of NS equation on the other -- whether it is indeed the nonrelativistic limit of semiclassical Einstein equation. At least one form of the NS equation, that proposed by Diosi and concurred by Penrose,  we find it to be at odds with general relativity theory, as was explained in Sec. 5.2.2.

\subsection{Level 1: Relativistic Matter Fields in Strong Gravitational Field. Einstein-Klein-Gordon Equation}

We now enter the relativistic realm, both for the quantum field and for gravity: Schr\"odinger equation is upgraded to Klein-Gordon equation for scalar particles or Dirac equation for spinors. The effect of curvature enters in the wave equation for a scalar field through the Laplace-Beltrami operator $\nabla^2 \Phi - m^2 \Phi = 0$. This theory has been applied to treat self-gravitating particles \cite{Ruffini} or boson stars \cite{Guzman}. Looking for a solution of the metric tensor sourced by the relativistic field requires that the Einstein equation and the KG equation be solved simultaneously in a self-consistent manner, namely, $G_{\mu\nu} = 8 \pi G T_{\mu\nu}(\Phi)$ where $ G_{\mu\nu}$ is the Einstein tensor and $T_{\mu\nu}(\Phi) $ is the stress energy tensor of the scalar field. Note at this level the field can be viewed either as  first quantized or second quantized. As a first quantized field  Guzman et al have shown that the E-KG equation reduces in the non-relativistic, weak field limit to the NS equation.  Spherically symmetric solutions of the NS equation have been found by e.g., Moroz et al \cite{Tod}.

At the second quantized level  truly quantum field theoretical effects like particle creation from the vacuum begin to show up. At the `test field' level  in which a quantum field propagates in a fixed given curved spacetime, this is the realm of quantum field theory in curved spacetime (QFTCST) \cite{BirDav,ParTom}.  How this field affects the background spacetime is described by the semiclassical Einstein equation. One now enters the realm of relativistic semiclassical gravity.

\subsection{Level 2: Relativistic Semiclassical Gravity (RSCG): Semiclassical Einstein Equation (SCE), including Backreaction of Quantum Matter Field}

%M\/oller and Rosenfeld \cite{MolRos} in the 60s first discussed the union of classical gravity with quantum fields in what is known as semiclassical gravity (SCG) today.

Semiclassical gravity (SCG) has been used for a wide range of theories where gravity is treated classically and the matter field quantum mechanically, including the nonrelativistic NS equation. But the treatment of matter varies from one particle to many-particle systems to quantum fields, and even for quantum, there is a difference between first quantized and second quantized.  To avoid confusion we add the word relativistic to SCG to  refer to fully relativistic Einstein's theory for gravity valid under strong field conditions,  and fully relativistic quantum  fields at the second quantized level for matter.  One known example of such a theory is that based on the semiclassical Einstein equation: $G_{\mu\nu} = 8 \pi G <T_{\mu\nu}(\Phi)>$ where $T_{\mu\nu}(\Phi) $ is the stress energy tensor of the matter field, here represented by a scalar field $\Phi$, and $< >$ denotes taking the expectation value with respect to a certain quantum state.

Note in general because of the sum over all modes there is ultraviolet divergence in this expression. Much of the effort in the field of QFTCST
%(which began in the late 60s with particle creation studies, epitomized in Hawking's effect in 74)
in the mid-70s focused on finding ways to regularize or renormalize these divergences. By 1978 when the results obtained  by different regularization approaches more or less converged serious studies of RSCG began, under the theme of "backreaction problems" which went on to the 80's -- e.g., the backreaction of vacuum energy of quantum fields (such as the Casimir effect) and  particle creation (from the vacuum) on the dynamics of the spacetime. This requires a self consistent solution of both the semiclassical Einstein equation governing the spacetime dynamics and the quantum matter field equation.
%This was made possible only after investigations of the renormalization or regularization of the stress energy tensor

From physical considerations, the backreaction of quantum fields brings forth dissipation in the dynamics of spacetime through the SCE Eq. How to reckon the appearance of non-unitary terms in an otherwise unitary evolution dictated by Einstein's equation was the first conceptual challenge. Understanding this issue from the open quantum system viewpoint  was helpful in discovering the next level of structure, in the Einstein-Langevin equation.

\subsection{Level 3: Stochastic SC Gravity (SSCG): Einstein-Langevin Equation, including Fluctuations in Quantum Field and Metric}
Including fluctuations of quantum field as a source driving the semiclassical Einstein equation faces another challenge. Why and how should a noise term appear in the SCE equation?  These two issues were resolved by borrowing concepts in nonequilibrium statistical mechanics, namely, the existence of fluctuation-dissipation relations and the use of the (Feynman-Vernon) influence functional formalism to provide an analytic basis for the description of quantum noise. This was how semiclassical stochastic gravity theory came into being \cite{HuVerLivRev}. Further proof by Verdaguer et al \cite{MarVer} that the noise can be written in a covariant form and satisfies the divergence-free condition ensures its rightful place in the Einstein-Langevin equation \cite{ELE}.
%This is the theoretical platform for the investigation of problems described in the last part of this report.

%%%%%%%%%%%%%%%%%%%%%%%%%%

\section{Newton-Schr\"odinger Equation and Semiclassical Gravity}

The Newton-Schr\"odinger (NS) equations play a prominent role in alternative quantum theories (AQT)\cite{BassiRMP}, emergent quantum mechanics \cite{AdlerCUP}, macroscopic quantum mechanics \cite{ChenMQM}, gravitational decoherence \cite{AHmasteq,Blen13}(as in the Diosi-Penrose models \cite{Diosi,Penrose}) and semiclassical gravity \cite{HuVerLivRev}.  The class of theories built upon these equations have drawn increasing attention because experimentalists often use it as the conceptual framework and technical platform for understanding the interaction of quantum matter with classical gravity and to compare their prospective laboratory results.
It is thus timely and necessary to  explore the assumptions entering into the construction of these equations and the soundness of the theories built upon them, especially in their relations to general relativity (GR) and quantum field theory (QFT), the two well-tested theories governing the dynamics of classical spacetimes and quantum matter.

Since NS are often simplistically conjured as the weak field (WF) limit of GR and the nonrelativistic (NR) limit of QFT, their viability is usually conveniently assumed by proxy, courtesy their well-accepted progenitor theories. We are not convinced of this. In a recent paper \cite{AH-NS} Anastopoulos and I show that NSEs do not follow from general relativity (GR) and quantum field theory (QFT), and there are no `many-particle' NSEs.  We come to this conclusion from two considerations: 1) Working out a model (see \cite{AHmasteq}) with matter described by a  scalar field interacting with weak gravity, %solve the constraint, canonically quantize the system then take the nonrelativistic (NR) limit. This procedure is the
with a procedure the same as in deriving the NR limit of quantum electrodynamics (QED). %Let's call this the QED route.
2) Taking the NR limit of the semiclassical Einstein equation (SCE), %where matter is described by a quantum field and its expectation value with respect with some quantum state acts as a source in driving the Einstein equation. Let's call this the SCE route. SCE Eq is
the central equation of relativistic semiclassical gravity (RSCG) (see last section for the four levels of SCG), %\footnote{There are four levels of semiclassical gravity theories \cite{HuEmQM} and one needs be careful which level one refers to when debating issues, better use the  most developed level \cite{HuVerLivRev}.}
a fully covariant theory based on GR+QFT with self-consistent backreaction of quantum matter on the spacetime dynamics \cite{HuVerLivRev}. The key points are summarized in \cite{AH-NS1}.

%As stated earlier the best accepted theory which describes quantum matter in a classical spacetime based on quantum field theory and general relativity is relativistic semiclassical gravity (RSCG).
%It is a fully covariant relativistic theory constructed in the 80s based on quantum field theory in general curved spacetime with backreaction from the quantum matter fields, and when field fluctuations are included, the upgraded stochastic gravity program \cite{HuVerLivRev}.

%Putting aside the various desirability of the NS equations (e.g., position-basis decoherence) it is necessary to examine their mathematical foundation and physical implications in relation to GR+QFT

%We point out what could go wrong if one stays at the quantum mechanical level (which is the one particle state of quantum field theory) in the derivation. The many-body NS equation obtained this way is not a physical representation of how quantum matter is coupled to classical gravity. Subtleties like this abound in marking the difference between a quantum mechanical versus a QFT treatment of matter in the presence of gravity or in curved spacetimes. One such point is exemplified in this note.

Before we explain the differences between theories based on NSE and those obtained from GR+QFT it may be useful to first highlight their differences in physical predictions:

\subsection{Problems with the Newton-Schr\"odinger Equations (NSE)}

We mention three aspects here.\\

A. In NSE the {\it gravitational self-energy}  defines non-linear terms in Schr\"odinger's equation.  In Diosi's theory \cite{Diosi}, the gravitational self-energy defines a stochastic term in the master equation.
With GR+QFT gravitational self-energy only contributes to mass renormalization, at least in the weak field (WF) limit.The Newtonian interaction term at the field level induces a divergent self-energy contribution to the single-particle Hamiltonian. It does not induce nonlinear terms to the Schr\"odinger equation for any number of particles.
%From the perspective of GR+QFT the NS equation {\em is not} the evolution equation for the single-particle wave function.\\

B. The {\it  one-particle NS equation}
%(Note it is long known \cite{HarHor} that RSCG is the theory obtained as the large N limit of N component quantum fields living in a curved spacetime. )  There are important differences:
appears as the Hartree approximation for $N$ particle states as $N \rightarrow \infty$. Consider the ansatz $|\Psi\rangle = |\chi\rangle \otimes |\chi \rangle \ldots \otimes|\chi \rangle$ for a $N$-particle system. At the limit $N \rightarrow \infty$ the generation of particle correlations in time is suppressed and one gets an equation which reduces to the NS equation for $\chi$ \cite{Alicki,AdlerTDG}\footnote{Note it is long known \cite{HarHor} that RSCG is the theory obtained as the large N limit of N component quantum fields living in a curved spacetime. Roura and Verdaguer \cite{RVlargeN} further showed that the next to leading order large N expansion produces stochastic semiclassical gravity.}.
. However, in the Hartree approximation, $\chi({\bf r})$ is \textit{not} the wave-function $\psi(\bfr)$ of a single particle, but a \textit{collective variable} that describes a system of $N$ particles under a mean field approximation. %The NS equation is like the quantum analogue of the classical Vlasov equation which is obtained.

C. A severe problem of the NSE when applied to a single-particle wave function is its {\it probabilistic interpretation}. Consider two statistical ensembles of particles one of which is described by the wave-function $\psi_1(\bfr)$ and the other by the wave function $\psi_2(\bfr)$. The ensemble obtained from mixing these ensembles with equal weight is described in standard quantum theory by the density matrix $\rho(\bfr, \bfr') = \frac{1}{2} [\psi_1(\bfr) \psi^*_1(\bfr') +   \psi_2(\bfr) \psi^*_2(\bfr')]$. The usual Schr\"odinger evolution guarantees that the probabilistic interpretation of the density matrix remains consistent under time evolution $\rho_t(\bfr,\bfr') = \frac{1}{2} [\psi_1(\bfr,t) \psi^*_1(\bfr',t) +  \frac{1}{2} [\psi_2(\bfr,t) \psi^*_2(\bfr',t)]$. This property does not apply for non-linear evolutions of the wave-functions.  The problem of nonlinearity in  quantum mechanics is an old issue which many AQTs are aware of, so we will just mention it here without further pursuit.

In what follows we will show that the only meaningful description of quantum matter interacting with classical gravity is if the matter degrees of freedom are described in terms of quantum fields, \textit{not} in terms of single-particle wave functions in quantum mechanics.

\subsection{NS Equation not from GR + QFT}

The NS equation governing the wave function of a single particle $\psi(\bfr,t)$ is of the form
\begin{eqnarray}
i \frac{\partial}{\partial t} \psi = - \frac{\hbar^2}{2m} \nabla^2 \psi + m^2 V_N[\psi]  \psi, \label{NS}
\end{eqnarray}
where %the potential energy $U(\bfr)$ from the gravitational interaction is of the form $U(\bfr)= m V_N(\bfr)$, where
$V_N(\bfr)$ is the (normalized) gravitational (Newtonian) potential given by
\begin{eqnarray}
V_N({\bf r},t) = - \int d{\bf r'} \frac{|\psi({\bf r'},t)|^2}{|{\bf r} - {\bf r'}|}.
\end{eqnarray}
It satisfies the Poisson equation
\begin{eqnarray}
\nabla^2 V_N = 4 \pi G \mu,
\label{Poisson}
\end{eqnarray}
where $\mu= m|\psi(\bfr,t)|^2 $ is the mass density, the nonrelativistic (slow motion) limit of energy density $\varepsilon = T_{00}$ (see below).

The Newton-Schr\"odinger equation predicts spatial localization of the wave-function, and decoherence only as a consequence of spatial localization. This ``collapse of the wave function" in space for macroscopic objects is a big `selling-point' of NS equations in many AQTs. Its desirable attributes aside, the logical foundation of the Newton-Schrodinger equation seems shaky to us. The naive identification of Newton as weak field limit of GR and Schrodinger equation as the nonrelativistic limit of QFT is likely behind the justification of NS equations. Here, one should exercise caution, as illustrated below:  E.g., on the GR side, not to identify gravitational potential as dynamical variables, and on the QFT side, not to mistake a field as a collection of particles described by single particle wave functions.
%For one, the coupling of quantum matter to gravity changes nature when approximations are introduced in the separate gravity (WF) and quantum (NR) sectors thus violating the covariance property which hold them together in RSCG \footnote{Recall at the inception of QFTCST even a simple concept like the vacuum merited  much careful considerations  because different vacua carry different physical meanings in curved spacetime. Gravitational self  energy is a subtle issue when quantum matter is involved.}.

\subsection{Non-relativistic weak field limit of SCE equation}

The central equation of relativistic semiclassical gravity (RSCG) is the semiclassical Einstein equation (SCE), and when quantum field fluctuations are included, the Einstein-Langevin equation, the centerpiece of stochastic semiclassical gravity \cite{HuVerLivRev}.  We examine the nonrelativistic limit of SCE and show that it is qualitatively different from the `many-particle' NS equation derived in \cite{ChenNS}.

The SCE Equation is in the form \footnote{We prefer calling this the semiclassical Einstein equation over the M\/oller-Rosenfeld equation because, after all, it is Einstein's equation with a quantum matter source.}
$G_{\mu \nu} = 8 \pi G \langle \Psi|\hat{T}_{\mu \nu}|\Psi \rangle$, where $\langle \hat{T}_{\mu \nu} \rangle$ is the expectation value of the stress energy density operator $\hat{T}_{\mu \nu}$ with respect to a given (Heisenberg-picture)  quantum state $| \Psi \rangle$ of the field.

In the weak field limit the spacetime metric has the form $ds^2 = (1 - 2 V)dt^2 - d{\bf r}^2$, and the non-relativistic limit of the semi-classical Einstein equation takes the form
\begin{eqnarray}
\nabla^2 V = 4 \pi G \langle \hat{\varepsilon}\rangle, \label{semi}
\end{eqnarray}
where $\hat{\varepsilon} = \hat{T}_{00}$ is the energy density operator. This can be solved to yield
\begin{eqnarray}
V(\bfr) = - G \int d{\bf r'} \frac{\langle \hat \varepsilon({\bf r'})\rangle}{|{\bf r} - {\bf r'}|}. \label{ehat}
\end{eqnarray}
The expectation value of the stress energy tensor in general has ultraviolet divergences and need be regularized. The procedures have been established since the mid-70's (see, e.g., \cite{BirDav}).

Two key differences between the NR limit of SCE and NSE are: i) the energy density $\hat \varepsilon(\bfr)$ is an operator, not a c-number. The Newtonian potential is not a dynamical object in GR, %just like the electric potential is not dynamical in QED,
but subject to constraint conditions.  ii) the state $|\Psi\rangle$ of a field is a  $N$-particle wave function. Quantum matter is coupled to classical gravity as a mean-field theory, well defined only when $N$ is sufficiently large.

The (misplaced) procedure  leading one from SCE to a NS equation  is the treatment of $ m |\psi(\bfr,t)|^2 $ as a mass density for a single particle, while in fact it is a quantum observable that corresponds to an operator $\hat{\epsilon}({\bf r}) = \hat{\psi}^{\dagger}({\bf r}) \hat{\psi}({\bf r})$ in the QFT Hilbert space when the matter degrees of freedom are treated as quantum fields $\hat{\psi}({\bf r})$ and $\hat{\psi}^{\dagger}({\bf r})$, as they need be. Not treating these quantities as operators bears the consequences A and B.

\subsection{Analog to the nonrelativistic limit of QED}

To cross check these observations we have carried out an independent calculation for matter described by a  scalar field interacting with weak gravity, following the same procedures laid out in \cite{AHmasteq}, namely, solve the constraint, canonically quantize the system, then take the nonrelativistic limit. This procedure is same as in obtaining  the non-relativistic limit of QED. We obtain the a Schr\"odinger equation
\begin{eqnarray}
i \partial |\psi \rangle / \partial t = \hat{H} |\psi \rangle,
\label{eqsch} \end{eqnarray}
with
\begin{eqnarray}
\hat{H} = - \frac{\hbar^2}{2m} \int d{\bf r}
\hat{\psi}^{\dagger}({\bf r}) \nabla^2 \hat{\psi}(\bfr) %\\ \nn
- G \int  \int d{\bf r} d{\bf r'} \frac{(\hat{\psi}^{\dagger}\hat{\psi})({\bf r}) (\hat{\psi}^{\dagger}\hat{\psi})({\bf r'})}{|{\bf r} - {\bf r'}|}. \label{ham}
\end{eqnarray}The electromagnetic analog of this equation with the Coulomb potential replacing the gravitational potential here is widely used in condensed matter physics (see \cite{AH-NS} for details).

The matrix elements of the operator (\ref{ham}) on the single-particle states $|\chi \rangle$ define the single-particle Hamiltonian:
\begin{eqnarray}
\langle \chi_2|\hat{H}|\chi_1 \rangle = - \frac{\hbar^2}{2m} \int d{\bf r} \chi_2^*({\bf r})\nabla^2 \chi_1({\bf r}) %\\ \nn
 - G \int d {\bf r} d{\bf r'} \frac{\chi_2^*({\bf r'})  \chi_1({\bf r}) \delta ({\bf r} - {\bf r'})}{|{\bf r} - {\bf r'}|}.
\end{eqnarray}
It is clear that Eq. (\ref{ham}) is very different from the NS equation (\ref{NS}) when considering a single particle state. % of the form (\ref{1par}).
For single-particle states the gravitational interaction leads only to a mass-renormalization term (similar to mass renormalization in QED). This is point A we made above. %The same mistake of treating a field as a product state of single particle wave function leads to the difference described in point B above.
Using the Hartree approximation to Eq. (4) leads to the same result as the NR WF limit of SCE, not NSE. This is Point B we made earlier. Details of this calculation are in \cite{AH-NS}.

Our analysis via two routes based on GR+QFT shows that NSEs are not derivable from them. Coupling of classical gravity with quantum matter can only be via mean fields. There are no $N$-particle NSEs. Theories based on Newton-Schr\"odinger equations assume unknown physics.

%%%%%%%%%%%%%%%%%%%%%%%%%%%%%%%%%%%%%%%%%%%%%%%%%%%%%%%%%%%%%%%%%%%%%%%%%%%%%%%%
%\vskip .5cm \centerline{\Large {\bf Part II.  Gravitational Decoherence: Master Equation for \\}}
%\centerline{\Large {\bf exploring the conflicts between Gravitation and Quantum}}
%\vskip .3cm
%%%%%%%%%%%%%%%%%%%%%%%%%%%%%%%%%%%%%%%%%%%%%%%%%%%%%%%%%%%%%%%%%%%%%%%%%%%%%%%
%%%%%%%%%%%%%%%%%%%%%%%%%%%%%%%%%%%%%%%%%%%%%%%%%%%%%%%%%%%%%%%%%%%%%%%%%%%%%%%%%%%%%%%%%%%%%%%%%%%%%%%%%%

\section{Gravitational Decoherence}
%\centerline{\Large {\bf Observations of interplay between Gravitation and Quantum}h}

%For a summary of our prior work on decoherence in quantum gravity, intrinsic decoherence and gravitational decoherence please refer to Introduction of \cite{AHmasterq}.

\subsection{Master Equations from GR + QFT: Our Analysis and Main Results}

%\noindent \textbf{Issues} With the meaning defined above, the questions we ask in this paper are.
%\begin{enumerate} \item Whether and how quantum matter is decohered by a gravitational field, under weak-field conditions. \item How gravitational decoherence differs from decoherence by a non-gravitational environment. \item What are the special features of gravitational decoherence-- the decoherence basis and rate. \end{enumerate}

%We find answers to these questions from a first-principles treatment of gravitational decoherence, viewing gravitational perturbations as an environment that affects the evolution of quantum particles. We employ general relativity (GR) for the description of gravity and quantum field theory (QFT) for the description of the matter degrees of freedom. Our derivations proceed from the general case to specific systems. We want to establish a general method for the study of gravitational decoherence that can be applied to many different physical situations. We try to avoid complex modeling assumptions, having as our main guide the mathematical structures of the two well-proven theories.

%\noindent \textbf{Our analysis: Open system of quantum matter with gravity induced effects} \\

%\textbf{Setup} The system under consideration is a massive scalar field, interacting with gravity described by classical general relativity. In the weak field limit, we describe gravitational perturbations in the linearized approximation.

The procedures we took in \cite{AHmasteq} are as follows:

First step: Start with the classical action of a massive scalar field interacting with gravity described by the Einstein-Hilbert action. Linearize the Einstein-Hilbert action around the Minkowski spacetime. Look at the weak-field regime. We do this for two reasons: a) we want results which can be tested in laboratory experiments at today's low energy  (in contrast to strong field conditions, as found in the early universe or late time black holes). b) In the derivation of the master equation for consideration of gravitational decoherence  the tracing-out of the gravitational field is not technically feasible, except for linearized gravitational perturbations.

The second step is to perform a 3+1 decomposition of the action and construct the associated Hamiltonian. Identify the constraints of the system and solve them at the classical level, expressing the Hamiltonian in term of the true physical degrees of freedom of the theory, namely, the transverse-traceless perturbations for gravity and the scalar field. The third step is to canonically quantize the scalar field and the gravitational perturbations together, to ensure the consistency between these two sectors from the beginning.

The fourth step  (after Eq. (21) of \cite{AHmasteq}) is to  trace over the gravitational field acting as its environment   to obtain a master equation for the reduced density matrix of the quantum matter field, including  the backreaction of the gravitational degrees of freedom. The system under consideration is formally similar to a quantum Brownian motion (QBM) model \cite{QBM,HPZ}.

The master equation for the reduced density matrix $\hat{\rho}_1$ of one non-relativistic quantum particle in 3D interacting with weak perturbative gravity, valid to first order in $\kappa= 8\pi G$, is given by
\begin{eqnarray}
\frac{\partial \hat{\rho}_1}{\partial t} = - \frac{i}{2m_R}
[\hat{{\bf p}}^2,\hat{\rho}_1]- \frac{ \kappa \Theta}{18m_R^2}
(\delta^{ij} \delta^{kl} + \delta^{ik}\delta^{jl}) [\hat{p}_i
\hat{p}_j,[\hat{p}_k\hat{p}_l, \hat{\rho}_1]] \label{me1}
\end{eqnarray}
where $p_i$ are the momentum components of the particle, $m_R$ is renormalized mass and $\Theta$ has meaning explained below.  This master equation enables gravitational decoherence studies and other related tasks. The fifth and final step is to project to a one particle state, then take the non-relativistic limit. We then use this nonrelativistic  master equation for the analysis of gravitational decoherence in  a single quantum particle. \\

%We point out the gauge nature of time and space reparameterizations in matter-gravity couplings, and warn that `intrinsic' decoherence or alternative quantum theories invoking stochastic dynamics arising from temporal or spatial fluctuations violate this fundamental symmetry of classical general relativity.

%After this we  follow the standard methodology of open quantum systems \cite{Dav, BP} in order to derive the 2nd order (perturbative) master equation for the quantum scalar field.

\noindent \textbf{Main Results}

\begin{enumerate}

\item
%We derive from accepted theories for quantum matter (QFT) and classical gravity (GR) a master equation describing a moving particle interacting with a weak gravitational field.

A special feature of decoherence by the gravitational field (in the non-relativistic limit) is the decoherence in the energy (momentum squared) basis, but not (directly) to decoherence in the position basis.  This is a direct challenge to theories (such as that proposed by Diosi and concurred by Penrose), which assume a potential energy term so that decoherence occurs in the position basis.  Our analysis shows that this class of theories violates the principles of general relativity.

%\item  We examine the significance of space and time reparameterizations in the description of a quantum field interacting with linearized gravity. We find that in order to obtain a Poincar\'e covariant description of the quantum field on Minkowski spacetime, it is {\em necessary} to fix the gauge. A gauge-invariant treatment of the associated constraints  does not appear compatible with the structures of Poincar\'e covariant QFT. This is the physical rationale for Penrose's gravity-induced decoherence \cite{Penrose}, putting aside the form of the master equation, as we have remarked previously.

\item Many approaches to gravitational or fundamental decoherence proceed by modeling temporal or spatial fluctuations in terms of  stochastic processes. However, such fluctuations correspond to time or space reparameterizations, which are pure gauge variables, with no dynamical content, according to classical GR.
     The assignment of dynamical content to such reparameterizations implicitly presupposes an underlying theory that violates the fundamental symmetry of classical GR.

\item The decoherence rate depends not only on the matter-gravity coupling, but also on the intrinsic properties of the environment,  such as its spectral density which reflects to some extent the characteristics of its sub-constituents composition. Measurement of the gravitational decoherence rate, if this effect due to gravity can be cleanly separated from other sources, may provide valuable information about the statistical properties of the sub-constituents, or what we called the ``textures", of spacetime.

\end{enumerate}

\section{Constraining Alternative Quantum Theories (AQT)}

\noindent{\bf Diosi's theory-- Version D} \cite{Diosi}\\
%is based on the Newton-Schr\"odinger or the Newton- von Neumann equation.

Diosi proposed a  master equation of the form
\begin{eqnarray}
\frac{\partial \hat{\rho}}{\partial t} = - i [\hat{H}, \hat{\rho}] - \frac{1}{4} \kappa G \int d {\bf r}_1 d {\bf r}_2 [ \hat{\mu}({\bf r}_1),[\hat{\mu}({\bf r}_2), \hat{\rho}]] \frac{1}{|{\bf r}_1 - {\bf r}_2|}.
\end{eqnarray}
where $\hat{\mu}({\bf r})$ is the mass density operator for the system and $\kappa$ a constant of order unity.
Diosi's master equation predicts decoherence of superpositions of macroscopically distinct states $X$ and $Y$ with a decoherence time $\tau_{dec} = {2 \hbar}/[{2U_D(X,Y) - U_D(X, X) - U_D(Y,Y)}]$, where
\begin{eqnarray}
U_D(X, Y) = - G \int d {\bf r}_1 d {\bf r}_2 \frac{f({\bf r}_1; X)f({\bf r}_2; Y)}{|{\bf r}_1 - {\bf r}_2|},
\end{eqnarray}
with $X$ and $Y$ parameterizing the distributions $\mu$. Typically one thinks of $X$ and $Y$ as centers of mass, whence the theory predicts decoherence in position.

One consequence of our investigation is the observation that Diosi's master equation cannot be derived from the framework of GR+QFT. It comes from the following considerations:
%\noindent \textbf{Gravitational decoherence in theories derivable from GR+QFT vs those not derivable}
General Relativity implies that the Newtonian interaction follows from the theory's Hamiltonian constraint. The solution of the constraint leads to a modification of the Hamiltonian through the addition of a Newtonian interaction term (in the non-relativistic limit):    $    H = H_0 - G \int d{\bf r}_1 d {\bf r}_2 \frac{f({\bf r}_1) f({\bf r}_2)}{|{\bf r}_1 - {\bf r}_2|}.$ Hence, the consistent quantization of the theory   should place the Newtonian interaction term as a part of the quantum {\em Hamiltonian}, not as part of the non-unitary dynamics.
There is no reason to structure the postulated non-unitary terms as a Newtonian interaction term, as is the case in Diosi's master equation.  This is forced upon as an assumption which contradicts general relativity.
%It is not in any sense a consequence of GR and QFT when applied to gravitational quantum physics.

More generally, we feel that general relativity suggests a very different class of fundamental decoherence models with different reduction basis and type of noise.  Working this out explicitly can make the comparison with the alternative models more quantifiable.\\

\noindent{\bf Other alternative theories} of quantum mechanics `aided' (or `interceded')  by gravitational effects -- at low energy (in contrast to the Planck scale) include the so called `continuous spontaneous localization' (CSL) models of Girahdi-Remini-Weber (GRW)- Pearle  \cite{GRWP} (see also work of Bassi et al \cite{BassiRMP}). The state reduction in these schemes is often facilitated by considering stochastic processes on the quantum system's Hilbert space and stochastic Schr\"odinger equations are often suggested as an alternative to quantum mechanics.
%Despite good arguments provided for this class of theories,
We will not address them here because the source of noise is phenomenologically motivated. We focus in the above on the Diosi-Penrose theories because it highlights the conflicts between gravity with quantum in a more transparent way - even if finally proven wrong, either way.

%\subsubsection{RSCG and SSCG: Well-founded theories for exploring issues of concern to AQTs}

\section{Role of Gravity in Macroscopic Quantum Phenomena}

Historically a primary motivation for introducing the continuous collapse models (CLS) is trying to make sense of the readily collapsed wave function of macroscopic objects while preserving the wave function in the microscopic realm. There are two main features of this class of models. They are (from \cite{BassiRMP} p. 482): 1) nonlinearity,  2) stochasticity. There are also two requirements:   3) no superluminal signaling -- this is forbidden from the start in RSCG since the basic principles of quantum field theory and relativity are observed.  4) an amplification mechanism -- an important issue which we feel needs more investigation. This is one aim of Chen's program on macroscopic quantum mechanics (MQM) \cite{ChenMQM} and our work on macroscopic quantum phenomena \cite{MQP1,MQP2,DICE12,MQP3}.  We now turn to this issue.

The main motivation  of the recent theoretical work by Chen's group \cite{ChenNS} is to adopt the NS equation for many particles (proposed by Diosi and others) to macroscopic objects and find out the differences from predictions of QM. They looked into the interaction between particles, the separation of scales in the dynamics of the center of mass variable from other variables. On this specific issue, a similar question was raised earlier by Chou Hu and Yu \cite{CHY} who set out to find the conditions where a ``Center of Mass Axiom" is observed. The key finding of CHY is, for interaction potentials dependent only on the separation between any two oscillators, the master equation for N oscillators has the same form as the HPZ master equation \cite{HPZ} for a single oscillator.  Studying N oscillator systems will enable one to see how their interactions affect the quantum features of the macroscopic object they form.

As for gravitational decoherence the master equation derived recently \cite{AHmasteq} applies to configurations with any number of particles. This is because a master equation for quantum matter fields was first derived before projecting it to the single-particle subspace. One can easily project it to any particle number state to obtain a master equation for N particles. It would be interesting to combine these two strands of investigations to see how gravity affects a quantum composite system from micro to meso to macro scales. We emphasize again the fundamental differences between our results which are based on GR+ QFT and those based on the (many-particle) NS equations which are not \cite{AH-NS1}.

%The latter simplifies significantly in the non-relativistic regime, and allows for the determination of the decoherence rate.

%\section{Gravity's Role in Micro - Macro / Quantum - Classical Issues}

\section{Observing Gravitational Decoherence -- Contributing Factors}

%\subsection{Fundamental Theoretical Issues}

%There are several subtle yet important points about gravitational decoherence which are not well appreciated or  recognized in the literature. One is that the Newtonian force term always appears in the Hamiltonian part which gives unitary evolution. \textit{This term which contains Newton's constant does not lead to decoherence}.  Another point is \textit{the pure gauge nature of space-time fluctuation-induced decoherence} -- decoherence due to uncertainties or  fluctuations in the position or temporal coordinates of physical events -- in  a large class of so-called intrinsic or fundamental decoherence models \cite{IntDec}. These theories describe such fluctuations by classical stochastic processes. However, fluctuations in the time or space coordinates of an event are indistinguishable mathematically from time and space reparameterizations of the system. Viewing such reparameterizations as stochastic does not alter its gauge nature. The invariance of the theory under space and time reparameterizations follows from the diffeomorphism invariance of the classical action, a fundamental symmetry of general relativity. Thus, as pointed out by us in the recent paper \cite{AHgravdec} STI decoherence is unphysical. theories violate the fundamental symmetry of classical general relativity.

A major factor in the surge of attention paid to gravitational decoherence is because several well respected experimental groups showed interest in the measurement of such effects \cite{GravDecExpts}. This is corroborated by some theorists' claims that their predicted values are close to current observable experimental precision levels.  It is thus important to examine carefully the assumptions made in theories which assert a significant effect in gravity's power to decohere a quantum particle. For example, in  the recent paper of Blencowe \cite{Blen13} (which we referred to as Version B), two assumptions were made,  as follows:

\subsection{Is thermal graviton bath a tenable assumption?}

How effectively can gravity decohere a quantum object depends strongly on the assumptions about the nature of the gravitational field acting as an environment to the quantum system of interest. The usual assumption that Minkowski spacetime is the ground state of quantum gravity would imply that gravitational perturbations are very weak. If these are the source of gravitational decoherence then the effect is very weak. This is the result we obtained in \cite{AHmasteq}.  However, if general relativity is a hydrodynamic theory and gravity is in the nature of thermodynamics, Minkowski spacetime could presumably be identified with a macrostate (i.e., a coarse-grained state of the micro-structures). In this case, gravity acting like a thermal bath may serve efficiently as an agent of decoherence.  (This information is contained  in the $\Theta$ parameter in \cite{AHmasteq}, in the former case $\Theta = 0$, in the latter case it can be large.)

Taking the former and more traditional view, one needs to ask how easily gravitons are thermalized, and if so, what is the graviton bath temperature? The source of gravitons received in the experiment's immediate environment can either be astrophysical or as remnants from the early universe, which is more diffuse and of a stochastic nature.
For gravitons treated as quantized weak perturbations off Minkowski space, which provide the lowest common denominator for a gravitational source in the consideration of decoherence, one can take the graviton scattering processes (see e.g., Papini's review \cite{Papini}), both amongst themselves and gravitons with other particles or massive objects,  and calculate the cross sections of these processes. To get some idea of the conditions for thermalization, comparison with earlier results for $\lambda \phi^4$ theory \cite{phi4thermal} under weak coupling, and for non-Abelian theories in heavy ion collision and quark gluon plasma processes \cite{Yaffe}, under strong coupling, for the two limiting conditions, may be helpful.  

For gravitons of cosmological origin, Blencowe took the value of 1 degree K from \cite{KolTur}. We don't know how that value came about. The likely pathway of argument is to draw an analogy with neutrinos, e.g., from \cite{Weinberg}. Neutrinos have been detected, but not gravitons, they are a lot trickier\footnote{Dyson's lecture \cite{Dyson12} addresses these points, with a lucid and crispy description of the relevant issues.}. Because of the extremely weak nature of their interactions, we don't think graviton thermalization comes easily.  Referring to ``thermal gravitons" or `graviton bath" with a certain temperature requires more careful deliberation.

%One needs to take a closer look at both of these issues raised in our recent paper because there is a deeper consequence on this point. As explained in \cite{AHmasteq}, the derivation of the master equation requires the specification of an initial state (at the initial time in the dynamics of the quantum particle) of gravitational perturbations.  Different assumptions about the nature of gravitational fluctuations will be reflected in the choice of this initial state, parametrized by a phenomenological noise-temperature $\Theta$.  In effect, $\Theta$ contains information  like the ``temperature" and the ``spectral density" of a harmonic oscillator bath in the quantum Brownian model analogy. This underlying ``textures" of spacetime may contain information on the constituency of spacetime and thus the nature of gravity, be it fundamental (elemental) or effective (emergent) \cite{E/QG}.

%Thus, observation of the magnitude and features of gravitational decoherence may reveal the nature of gravity, whether it is elemental or emergent.

A somewhat equivalent way to look at this issue is the difference between (in the classical view) a superposition of gravitational waves and (in the quantum view) a mixed state of such superpositions. The estimate (in our view, over-estimate) of gravitational decoherence due to gravitational waves in the solar system (e.g., \cite{Reynaud}) bears on this essential point which can be the source of confusion and need be more clearly expounded (see, e.g., \cite{GravDecAP}).

%even the relaxation time of would suffice (e.g., via Boltzmann equation) , if one finds the proper correspondences.

\subsection{Does simple scaling up of quantum attributes apply to macroscopic objects?}

In addition to the assumption of a thermal bath for gravitons the other main reason why Blencowe obtained a large number (compared to ours) for the gravitationally induced decoherence rate is because he uses a simple scaling from a quantum particle to a massive object. For an initial superposition of ground and excited states of a single atom  the decoherence rate from his formula is $\sim 10^{-45} / sec$. This small rate (meaning it takes a very long time) is commensurate with our claim that for a weak gravitational perturbations background (at zero temperature) there is essentially no gravitational decoherence effect.  However, Blencowe continues, ``For a matter system comprising an Avogadro¡¯s number of atoms $\sim 1$ gram in a quantum superposition where all of the atoms are either in their ground state or all in their excited state," he got a decoherence rate of $\sim 10^2 $/sec. ``For a system with mass $\sim 1$ kg in such a superposition state, the gravitationally induced decoherence rate  projects is $\sim 10^8 $/sec. We believe Blencowe made an implicit assumption in how a macroscopic system's quantum behavior is directly related to the quantum features of its microscopic constituents.  This is a largely unexplored topic
under the general subject of macroscopic quantum phenomena (MQP).
This issue needs to be addressed before one can assuringly take the results for micro quantum objects and scale it up to macro domains. Interaction strength and quantum coherence amongst the sub-constituents at different levels of structure are expected to play an essential role (see, e.g., \cite{MQP3}), factors which in a fuller investigation may invalidate the projection of significant gravitational decoherence.

\vskip .5cm

\noindent {\bf Acknowledgment} The principal organizers, Gerhard Gr\"ossing and Jan Walleczek,  are to be thanked for making this meeting on foundational issues of physics possible, even lavishly so.  The main parts of this talk are based on two recent papers written with Charis Anastopoulos.

\bigskip

%\newpage
%%%%%%%%%%%
%\newpage

%\setcounter{page}{1}

\end{document}